# Self-decoupled tetrapodal perylene molecules for luminescence studies of isolated emitters on Au(111)

Talha Ijaz,[1] Ben Yang,[1] Ruipu Wang,[1] Jiazhe Zhu,[1] Aftab Farrukh,[1] Gong Chen,[1,2] Gregory Franc,[3] Yang Zhang,[1,a)] Andre Gourdon,[3,a)] Zhenchao Dong[1,a)]

[1]*Hefei National Laboratory for Physical Sciences at the Microscale, Department of Chemical Physics, University of Science and Technology of China, Hefei, Anhui 230026, China*

[2]*School of Physics and Engineering, Zhengzhou University, Zhengzhou 450052, China*

[3]*Centre d'Elaboration de Matériaux et d'Etudes Structurales, CNRS 29 rue Jeanne Marvig, F-31055 Toulouse, France*

[a)] **Authors to whom correspondence should be addressed:** zcdong@ustc.edu.cn, andre.gourdon@cemes.fr, and zhyangnano@ustc.edu.cn

**ABSTRACT**

Self-decoupled tetrapodal perylene molecules were designed, synthesized and deposited on the Au(111) surface through electrospray ionization technique. Photoluminescence and lifetime measurements show that the chromophore groups of the designed molecules are well decoupled from the gold substrate. Molecule-specific emissions from both neutral and anionic molecules were observed in the preliminary electroluminescence measurements. The emergence of significant emission when the tip is positioned at the molecular center suggests that there is a considerable vertical component of the transition dipole of the designed molecule along the tip axial direction. Our results may open up a new route for the realization of nano light sources and plasmonic devices based on organic molecules.

Scanning tunneling microscope (STM) induced molecular emission offers unprecedented opportunities for gaining insights into single-molecule characteristics by exploring optoelectronic phenomena occurring at the single-molecule level.[1-15] A key issue to obtain single molecular electroluminescence in STM is to electronically decouple the molecular emitters from the metallic substrate since the fluorescence of the emitters directly adsorbed on the metal surface would be quenched due to the direct electron transfer between molecules and metal substrate. The most widely used approach to achieve electronic decoupling is to physically introduce an insulating spacer between the molecule and metal substrate, such as oxides,[1] halides,[3] thiol layers,[16] and molecular multilayers.[2] A less explored approach is the chemical route that involves chemical modifications by adding spacer groups to the emitting unit, which would allow studying a wide range of functionally tunable molecules directly adsorbed on metal substrates.[17-20] Indeed, the electronic self-decoupling effect was achieved previously by chemically adding tripodal anchors to the porphyrin chromophore.[18] However, a majority of porphyrin molecules with tripodal anchors still tend to adopt flat-lying configurations after deposition on a metal substrate and thus still suffer from severe fluorescence quenching.[18] Consequently, how to further optimize the chemical design of anchor supported molecules to achieve efficient decoupling, regardless of the adsorption configurations, remains to be a challenging issue to be solved. One other challenge is how to align the transition dipole of the emitters along the vertical direction that would favor strong plasmonic enhancement? In this work, we shall first present the chemical design, synthesis, and deposition of self-decoupled tetrapodal perylene (TP) molecules, which is better optimized for self-decoupling properties. Then, we shall investigate the optical properties of the TP molecules on Au(111) using photoluminescence (PL) and scanning tunneling microscope induced luminescence(STML) techniques.

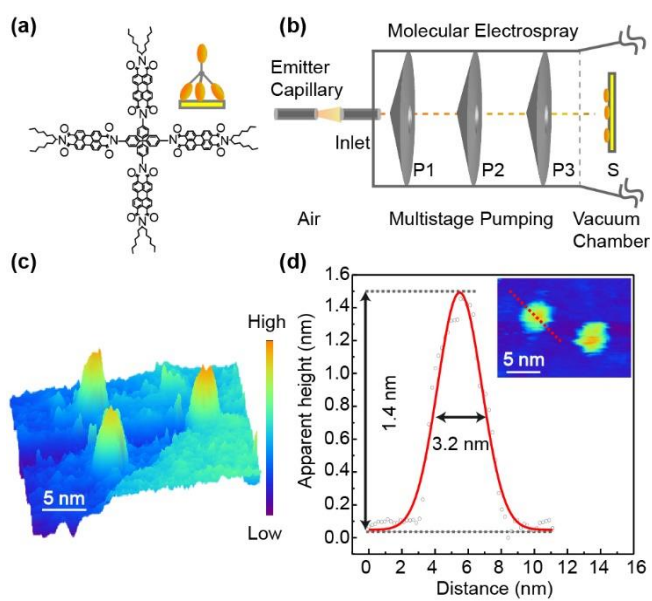

**FIG. 1.** (a) Chemical structure of the designed TP molecule, with the inset on the top-right showing the decoupling strategy. (b) Schematic of electrospray ionization for molecular deposition in vacuum. P1, P2, P3 shows differentially pumping stages and S represents the Au substrate. (c) A 3D view of STM topograph of single TP molecules on Au(111) (2.7 V, 2 pA, 27 × 22 nm$^2$). (d) Typical height profile of a single self-decoupled TP molecule on Au(111) for the line trace shown in the inset STM image (2.7 V, 2 pA, 20 × 15 nm$^2$).

Figure 1(a) shows the chemical structure of our designed molecule. The tetrahedral geometry is adopted to explore the idea that no matter in what configuration molecules are adsorbed on the surface there would always be one emitter pointing in the upward direction, as shown in the inset of Fig. 1(a). To this end, the tetraphenylmethane is selected to act as a tetrahedral backbone for the targeted tetrapodal molecule. Perylene diimide (PDI) molecules are chosen due to their high photochemical stability and quantum yield. PDI functions as both emitter and spacer, though separately, while alkyl chains (i.e., two hexylheptyl chains for each PDI unit) serve as solubilizing moieties and anchoring groups.

The target TP molecule, namely, tetra-4-[N-(1-Hexylheptyl)perylene-3,4:9,10-tetracarboxylicbisimid-N′-yl]phenylmethane, was obtained by a route similar to previous reference,[21] but by direct condensation of four PDI precursors with tetrakis(4-aminophenyl)methane in presence of zinc acetate/imidazole rather than by conversion to the tetraformamide. The deposition of isolated TP molecules on metal surfaces is challenging because of its large molecular weight and low thermal stability, making it impossible to transfer by thermal sublimation under ultrahigh vacuum (UHV) conditions. In order to overcome such difficulty, we employ the electrospray ionization (ESI) technique to deposit these molecules onto the Au(111) metal surface.[22,23] As shown in Fig. 1(b), a commercial ESI system (Molecular Spray UHV4) with three-stage pumping is mounted to the loadlock chamber. A solution with a concentration of ~34 μM was prepared by dissolving TP molecules in solvents consisting of a mixture of dichloromethane and methanol with a volume ratio of 4: 1. This solution was passed through a stainless steel emitter capillary held at ~1.8 kV. This leads to the formation of spray outside the vacuum system. The spray enters the pre-vacuum chamber through an aperture, then passes through differentially pumped vacuum stages, and finally deposits onto atomically flat Au(111) surfaces.[22] The vacuum of the loadlock chamber varies slightly during the ESI deposition for 300 seconds, typically from a base pressure of ~$1.5 \times 10^{-6}$ Torr to ~$2.0 \times 10^{-6}$ Torr.

All of the STM, PL, and STML experiments were performed in the observation chamber of a low-temperature UHV-STM (Unisoku) with a base pressure of ~$8 \times 10^{-11}$ Torr at ~78 K or ~7 K, unless otherwise noted. STM imaging was performed at constant-current mode with sample biased. A gold film with a thickness of ~200 nm was prepared by thermal deposition of gold on a freshly cleaved mica surface. Before ESI deposition, the atomically flat Au(111) surface was achieved by cycles of sputtering and annealing. Electrochemically etched silver tips used in the experiments were cleaned by electron bombardment and argon ion sputtering, followed by tip indentations to achieve desired nanocavity plasmon (NCP) modes for resonance enhancement. The optical setups were detailed in previous references.[7,9,24] For lifetime measurements, a 532 nm pulsed laser with a 35 ps pulse width was used and synchronized with the time-correlated single-photon counting (TCSPC) setup.

Figure 1(c) shows the STM image of deposited molecular species on the Au(111) surface. Isolated bright spots, with a surface coverage of ∼4%, can be readily identified. Typical dimension parameters measured for these bright spots are $3.2\pm0.2$ nm in width and $1.4\pm0.1$ nm in apparent height, respectively. Based on theoretical calculations with semiempirical quantum mechanical PM6 method,[25] typical values for the configuration-optimized TP molecule are ∼3.0 nm in both width and height. The consistency between the measured width of the bright spots and the calculated width of a TP molecule suggests that each bright spot is likely to correspond to a single TP molecule, as subsequently further confirmed by PL and STML measurements. In addition, at low coverage, the TP molecules tend to remain as isolated single entities instead of forming molecular aggregates, suggesting that the TP molecules are anchored on the Au(111) surface after ESI deposition.

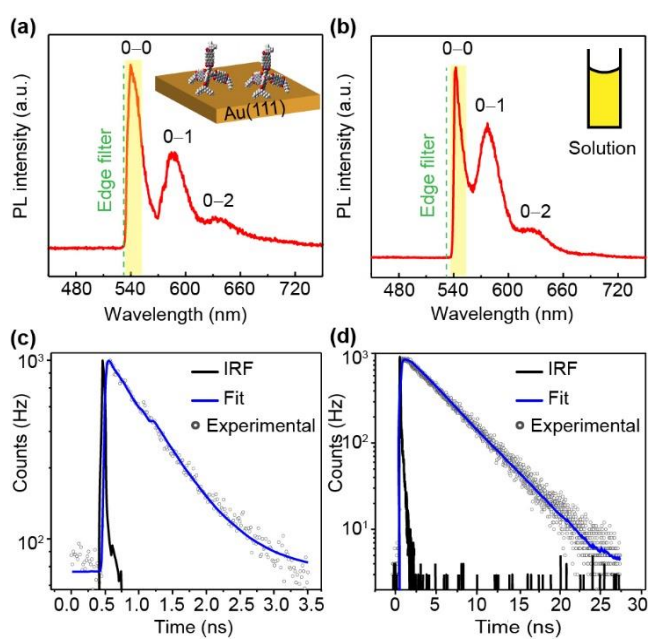

**FIG. 2.** (a) PL spectrum of self-decoupled TP molecules on Au(111) at 7 K, excited by a 532 nm pulse laser. (b) PL spectrum of TP molecules in a solution mixed with dichloromethane and methanol, measured at ambient temperature. (c) Time-resolved fluorescence decay curve of TP molecules adsorbed on Au(111) at 7 K. The blue curve is a fit convoluted with the instrument response function (IRF, black curve). (d) Time-resolved fluorescence decay curve of self-decoupled TP molecular solution measured at ambient temperature. A bandpass filter was used in the TCSPC lifetime measurements for the 0−0 emission.

In order to examine the self-decoupling efficiency of the designed TP molecules on Au(111), we first carried out photoluminescence measurements to reveal their emission feature. Figure 2(a) shows a typical PL spectrum acquired from the TP molecules adsorbed on Au(111), featuring three emission peaks at ~540 nm, ~585 nm and ~637 nm, respectively. These peak positions are similar to the spectrum acquired from TP molecules in solution, as shown in Fig. 2(b), and are also consistent with the emission peak values of PDI monomer molecules reported in the literature.[26-28] In addition, according to the mirror-image symmetry between the absorption and fluorescence spectra, the emission peak at ~540 nm can be assigned to the 0−0 line, and the peaks at ~585 nm and ~637 nm can be assigned to the associated vibronic progression with an energy difference of ~0.17 eV. The observation of clear TP molecular emission in PL suggests that the direct electron transfer between the emitter and the substrate is blocked and as a result, the emitter group (top PDI molecular unit) is indeed well decoupled from the underneath gold substrate. In other words, the chemical design strategy via a tetrahedral tetrapod works well.

In order to probe the influence of both the local environment around the emitter and the metal substrate on the exciton decay dynamics of TP molecules, we also performed fluorescence lifetime measurements on the 0−0 emission of the TP molecules adsorbed on Au(111) and in solution, respectively, as shown in Fig. 2(c) and 2(d). The fluorescence decay of the TP molecules in solution can be well fitted by a single exponential function with a lifetime of ~4.31(1) ns, which is very close to the PDI monomer lifetime of about ~4 ns reported in previous literature.[27,29] Such an agreement suggests that the tetropodal chemical design does not generate additional nonradiative channels that would otherwise quench the molecular fluorescence. However, the fluorescence of the TP molecule on Au(111) exhibits a biexponential decay feature with the major component (~83%) for a longer lifetime of ~0.58(2) ns and the minor component (~17%) for a shorter lifetime of ~0.12(1) ns. Both lifetimes are much shorter than that observed for the TP molecules in solution, which suggests that the presence of the metal substrate opens up fast energy transfer channels.[30,31] The shorter lifetime (~0.12 ns) might be assigned to the direct decay channel between the excited chromophore to the metal substrate through dipole-dipole interactions, while the relatively longer lifetime (~0.58 ns) might be attributed to the indirect decay channel that is associated with the interaction between the top perylene emitter and the underneath spacer perylene units, the latter are close to the metal substrate and could then dissipate the transferred energy rapidly to the substrate. In this context, the dominant role of the ~0.58 ns component, together with the observation of clear molecule-specific emission, indicates that such self-decoupled tetrapodal chemical design indeed suppresses effectively the fluorescence quenching from the metal substrate. All these PL results also give further justification that the bright individual spots observed in the STM image of Fig. 1(c) are indeed the target self-decoupled TP molecules.

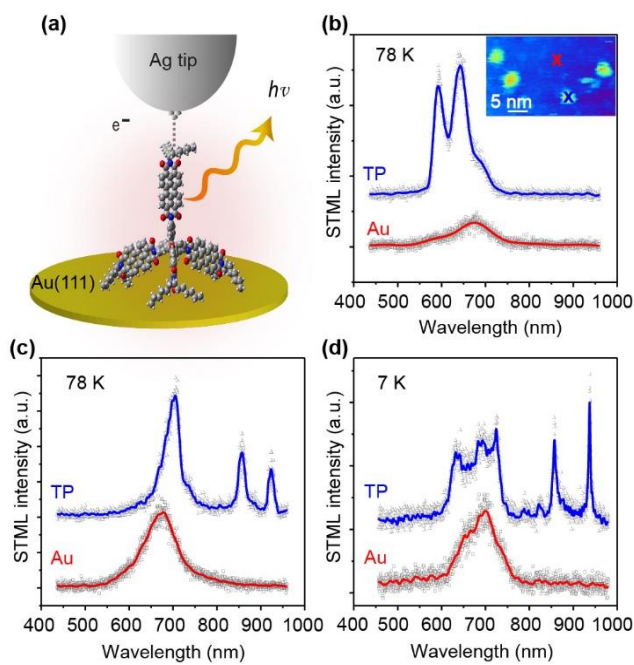

**FIG. 3.** (a) Schematic of STML setup for a single self-decoupled TP molecule on Au(111). (b) One example of STML spectra (2.7 V, 30 pA, 30 s) at 78 K measured at the labeled positions in the inset STM image (2.7 V, 2 pA, 33 × 20 nm$^2$). (c) Another example of STML spectra (2.7 V, 30 pA, 30 s) at 78 K. (d) One type of typical STML spectra (2.7 V, 2 pA, 300 s) acquired at 7 K. Solid lines show the smoothed curves with raw data denoted by the gray dots. The red curves are STML spectra measured on the bare Au(111) surface near TP molecules.

We also performed preliminary STML measurements on the TP molecules directly adsorbed on Au(111) to probe the electroluminescence behavior of individual TP molecules under tunneling electron excitations, as schematically illustrated in Fig. 3(a). Here, we would like to note that, unlike the constant observation of three characteristic emission peaks in the PL measurements for the TP molecules on Au(111) [Fig. 2(a)], STM induced luminescence acquired above individual TP molecules on Au(111) was found not only different from the PL spectrum, but also exhibiting a variety of different spectral profiles. Such distinctive luminescence behavior suggests that tunneling electrons (and tip approaching) can easily modify the molecular configuration or electronic states since TP is a complicated bulky molecule with long alkyl chains. The easily changing nature of the molecular configurations or electronic states under tunneling electron excitations makes it highly challenging to run

STML measurements and to understand their emission features as well. For example, as shown in Fig. 3(b), we observed two sharp emission peaks at ~593 nm and ~641 nm, respectively, which are evidently different from the broad NCP emission measured on the nearby bare Au(111) surface [red curve in Fig. 3(b)] and thus suggest the molecular origin of the emission. In addition, the observed double peaks are found to resemble the 0–1 and 0–2 vibronic features of the PL spectrum of TP on Au(111) with a similar energy spacing of ~0.16 eV despite small redshifts in the peak positions. Therefore, the double-peak STML spectrum is attributed to the fluorescence from a neutral TP molecule. The redshift of the peak positions in the STML measurement with respect to those in the PL measurements may originate from either configuration variations or the stark effect.[32-34] The reason for the absence of 0–0 emission is unclear, but might be related to the easily changing nature of molecular configurations upon electron excitations, which may favor vibronic emissions.

In addition to the double-peak spectral profile such as Fig. 3(b), STML spectra were often found to also exhibit sharp emission peaks in the long-wavelength region above 700 nm. As exemplified in Fig. 3(c), two sharp emission peaks at ~856 nm and ~923 nm were observed, which could be related to either anionic emissions from charged PDI units[35,36] or phosphorescence.[13,37] The relatively broad peak at ~700 nm might be related to the 0–3 emission of a neutral TP molecule in terms of the vibrational energy spacing of ~0.16 eV observed above, though the possibility stemming from molecule-modulated NCP emission cannot be completely excluded.[15,38] The observation of emissions from both neutral and charged TP molecules indicates that tunneling electrons not only can modify molecular configurations, but also can change molecular charging states.[11,13]

In an effort to improve the configurational stability of the TP molecules, we further cooled the system to the liquid-helium temperature (~7 K at the sample). However, the configurational stability of the TP molecules does not seem to improve since various types of STML spectral profiles were still observed at such a low temperature even for very small currents down to 2 pA, which suggests that the TP molecular configurations are very susceptible to perturbation by tunneling electrons. As shown in Fig 3(d), two very sharp emission peaks at ~856 nm and ~937 nm were observed, which can be attributed again to the anionic emission or phosphorescence of TP molecules. The origin of the broad emission around 700 nm with three minor peaks added on top (ranging from ~632 nm to ~725 nm) is still unclear but appears to contain mixed contributions from the TP molecule and NCP emission.

Notably, all the blue STML spectra shown in Fig. 3 were acquired when the tip was positioned above the center of individual TP molecules. The observation of significant molecule-specific emission intensity at the center position (stronger than or comparable to other molecular sites) is substantially different from the STML behavior on a flat-lying zinc phthalocyanine (ZnPc) molecule, where the intensity acquired at the ZnPc center is extremely weak because the ZnPc transition dipole is oriented horizontally and the associated total dipole symmetry of the whole tip–molecule–substrate system tends to cancel out.[4,7] Such a great contrast in the STML intensity at the molecular center between a TP and a ZnPc suggests that the transition dipole of the TP molecule on Au(111) is likely to stand up straight or at least contain considerable vertical component along the tip axial direction. Such a vertical-dipole configuration may offer new opportunities for probing possible strong coupling between the NCP and molecular emitter.

In summary, we have designed and synthesized a tetrahedron-like tetrapodal perylene molecule that can suppress fluorescence quenching from metal substrates through the built-in self-decoupling units, regardless of adsorption configurations. We have also demonstrated that molecular electrospray technique can be used to deposit such large functional but thermally unstable molecules on Au(111) for the luminescence study of isolated molecules. Both steady-state PL and transient lifetime measurements for isolated TP molecules on Au(111) justify the design concept through the observation of clear molecule-specific emissions and shortened lifetime components. Preliminary STML measurements on individual TP molecules on Au(111) indicate that tunneling electrons can readily modify the molecular configuration and even charging states, resulting in various luminescence spectral profiles and emission bands. The observation of significant emissions from the molecular center in STML measurements also suggests a standing-up orientation of the top-emitter transition dipole and further confirms the design concept. Future works should focus on the improvement of the stability of molecular configurations, so that molecular configurations, charging states, and luminescent properties can be better controlled and correlated. Our results provide new routes for controlling transition dipole orientations and for developing single-molecule light sources and organic optoelectronic devices.

This work is supported by the National Key R&D Program of China (grant numbers 2016YFA0200600 and 2017YFA0303500), the National Natural Science Foundation of China, the Chinese Academy of Sciences, and Anhui Initiative in Quantum Information Technologies. Talha Ijaz acknowledges support by the World Academy of Sciences. Yang Zhang acknowledges support by Excellent Young Scientist Foundation of the National Natural Science Foundation of China.